# Simulation Performance of MMSE Iterative Equalization with Soft Boolean Value Propagation

Aravindh Krishnamoorthy, Leela Srikar Muppirisetty, Ravi Jandial
ST-Ericsson (India) Private Limited
http://www.stericsson.com
{aravindh.k, srikar.ml, ravi.jandial}@stericsson.com

*Abstract*—The performance of MMSE Iterative Equalization based on MAP-SBVP and COD-MAP algorithms (for generating extrinsic information) are compared for fading and non-fading communication channels employing serial concatenated convolution codes.

MAP-SBVP is a convolution decoder using a conventional soft-MAP decoder followed by a soft-convolution encoder using the soft-boolean value propagation (SBVP).

From the simulations it is observed that for MMSE Iterative Equalization, MAP-SBVP performance is comparable to COD-MAP for fading and non-fading channels.

## I. INTRODUCTION

Iterative Equalization for Wireless Communication channel employing serial concatenated codes has been investigated in [5], [6] amongst others. A primary block in the Iterative Equalizer is the convolution decoder which generates the extrinsic information to be passed to the next iteration of Equalization.

COD-SOVA [4] and COD-MAP [3] are two well known algorithms of the convolution decoder used for this purpose.

24th November, 2011

## II. MAP-SBVP

MAP-SBVP is a combination of soft-MAP algorithm [3], soft-convolution encoder, and a hard-converter as shown in figure 1.

Soft-MAP algorithm provides the LLRs for output (uncoded) bits and these are converted into LLRs of input (coded) bits using the soft-convolution encoder and these serve as extrinsic information for the next iteration. In the last stage of iteration, the hard-outputs are taken as final decoded bits.

A similar scheme may also be setup for SOVA-SBVP in which SOVA algorithm [4] is used instead of soft-MAP.

### A. Soft-Boolean Value Propagation

Soft-boolean value propagation extends the boolean value propagation to LLRs [4]; and of particular interest is the $\oplus$ (XOR) operation which is used for convolution encoding. If $v_1$ and $v_2$ are two soft-values such that $v_1 = \lambda(b_1)$ is the LLR of bit $b_1$ and $v_2 = \lambda(b_2)$ is the LLR of bit $b_2$, then:

$$v_1 \boxplus v_2 = \lambda(b_1 \oplus b_2) \quad (1)$$
$$= \sigma(v_1)\,\sigma(v_2)\,\min(|v_1|, |v_2|) \quad (2)$$

Where $\sigma(x)$ is the 'sign' function given as

$$\sigma(x) = \begin{cases} -1 & \text{if } x < 0 \\ +1 & \text{if } x \geq 0 \end{cases}$$

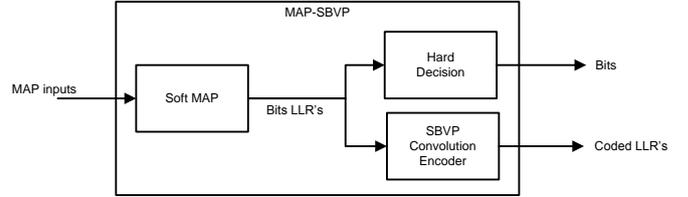

Fig. 1. MAP-SBVP

We refer to the $\boxplus$ operator as soft-XOR operator. Note that eqn. (2) is the approximate version of the soft-XOR operation as given in [4].

### B. Soft-Convolution Encoder

A soft-convolution encoder uses the $\boxplus$ (soft-XOR) operation instead of the $\oplus$ (XOR) operation. A rate 1/2 convolution encoder polynomial for bits and LLRs is given below.

| Convolution Code Polynomial (Bits) | Convolution Code Polynomial (LLR) |
|---|---|
| $1 \oplus D^3 \oplus D^4$ | $1 \boxplus D^3 \boxplus D^4$ |
| $1 \oplus D \oplus D^3 \oplus D^4$ | $1 \boxplus D \boxplus D^3 \boxplus D^4$ |

As an example, for an input LLR's bitstream

```
x = [-43.2565  -166.5584  12.5332 \
      28.7676  -114.6471  119.0915]
```

for the polynomial $1 \boxplus D^3 \boxplus D^4$ given above, the fifth output is computed as

$$\begin{aligned}
y(5) &= (1 \boxplus D^3 \boxplus D^4)\,x(5) \\
&= x(5) \boxplus x(2) \boxplus x(1) \\
&= \sigma(x(5))\,\sigma(x(2))\,\sigma(x(1))\,\min(|x(5)|, |x(2)|, |x(1)|) \\
&= -43.2565
\end{aligned}$$

## III. SIMULATION

The MATLAB based simulation testbench for Iterative Equalization [2] is used to simulate and compare the performance of COD-MAP and MAP-SBVP algorithms.

## A. Customization of the Simulation Testbench

The test-bench's [2] COD-MAP convolution decoder is modified to produce the LLRs of output bits as an additional output. These LLRs are then soft-convolution encoded as described in the sections above and used for equalization of subsequent iterations. The simulation parameters are as follows.

| Equalizer | Exact MMSE Equalizer (equ_exact_lin) |
|---|---|
| Channels | Channel (a), (b), (c) from [1] |
| Convolution Polynomials | K=5, Rate 1/2 |
| Puncturing and Interleaving | YES |

Channel (a) and (b) have good frequency-characteristics while channel (c) is highly frequency-selective.

## B. Simulation Results

Figures 2, 3, 4, 5 are the simulation BER results with no channel, channel (a), channel (b) and channel (c) respectively. Turbo-equalization has no advantage for the no-channel condition while minor improvements in BER performance are seen for fading channels. In general, the performance of MAP-SBVP is found to be comparable with COD-MAP.

## IV. CONCLUSION

The MAP-SBVP performance is compared against COD-MAP for MMSE Turbo Equalization and it is found that the performance of MAP-SBVP is comparable to COD-MAP for fading and non-fading channels.

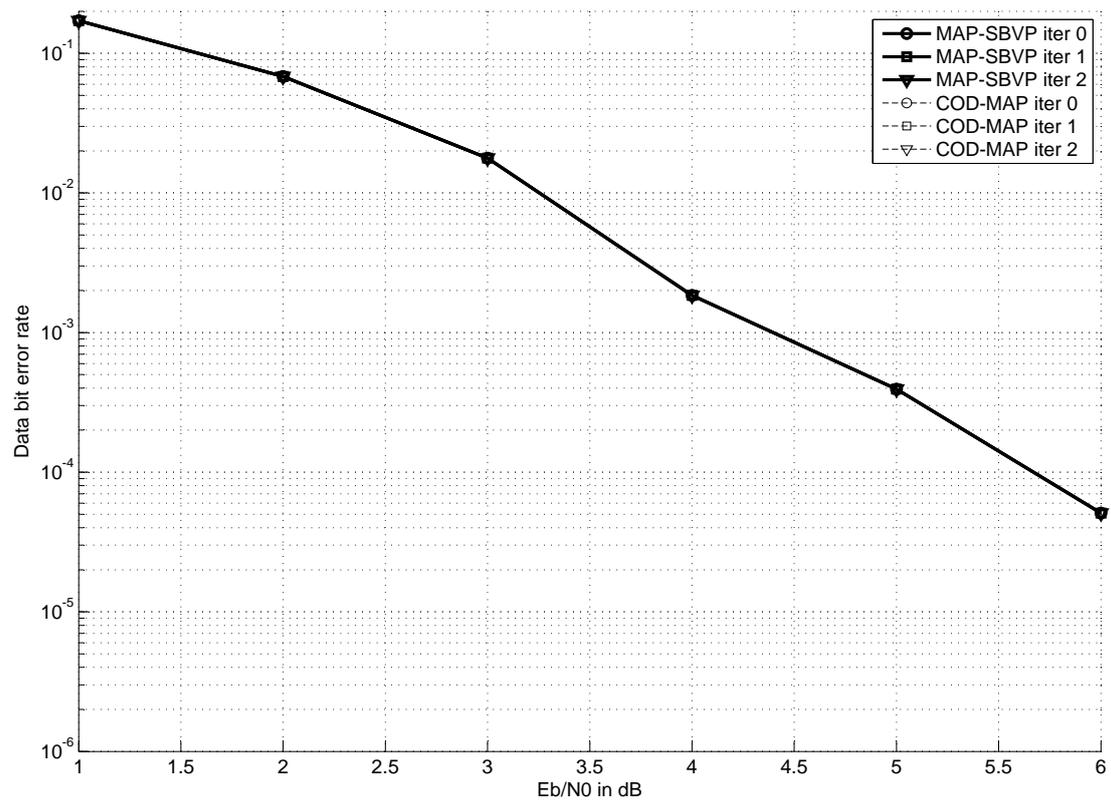

Fig. 2. No channel, QPSK Modulation, MMSE Equalization, Rate 1/2 (K=5) Convolution Coding and Puncturing, Block interleaving

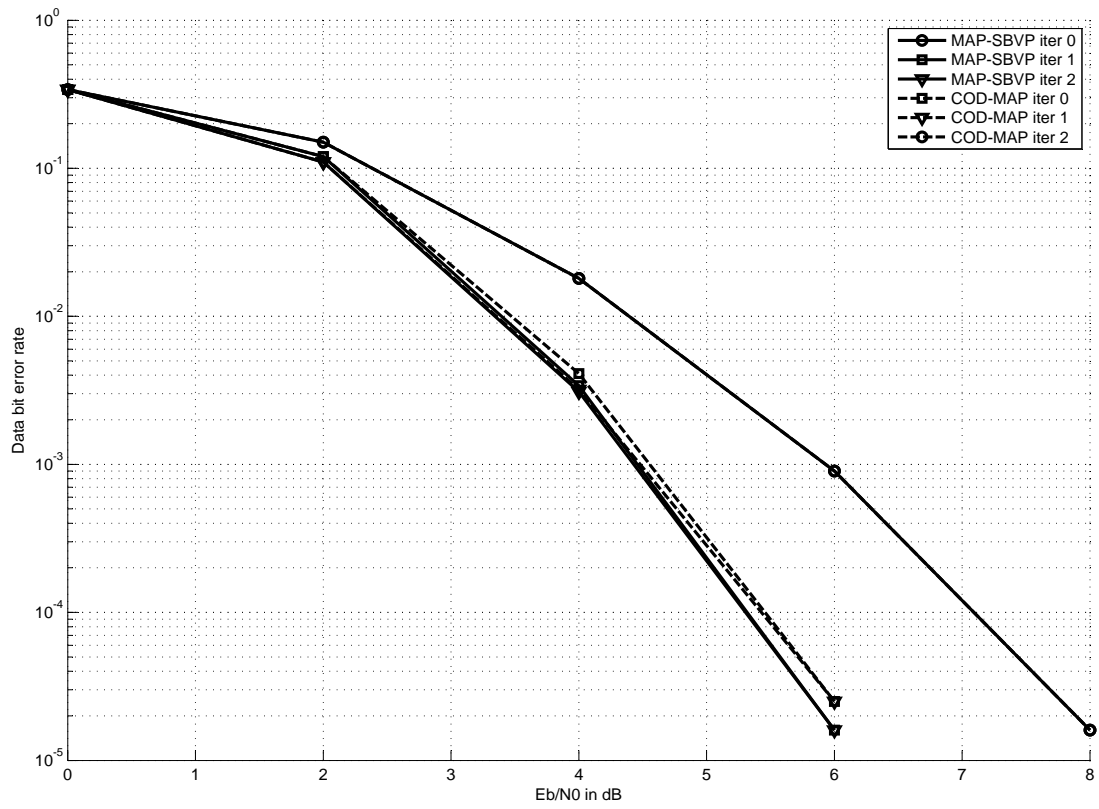

Fig. 3. Channel (a), QPSK Modulation, MMSE Equalization, Rate 1/2 (K=5) Convolution Coding and Puncturing, Block interleaving

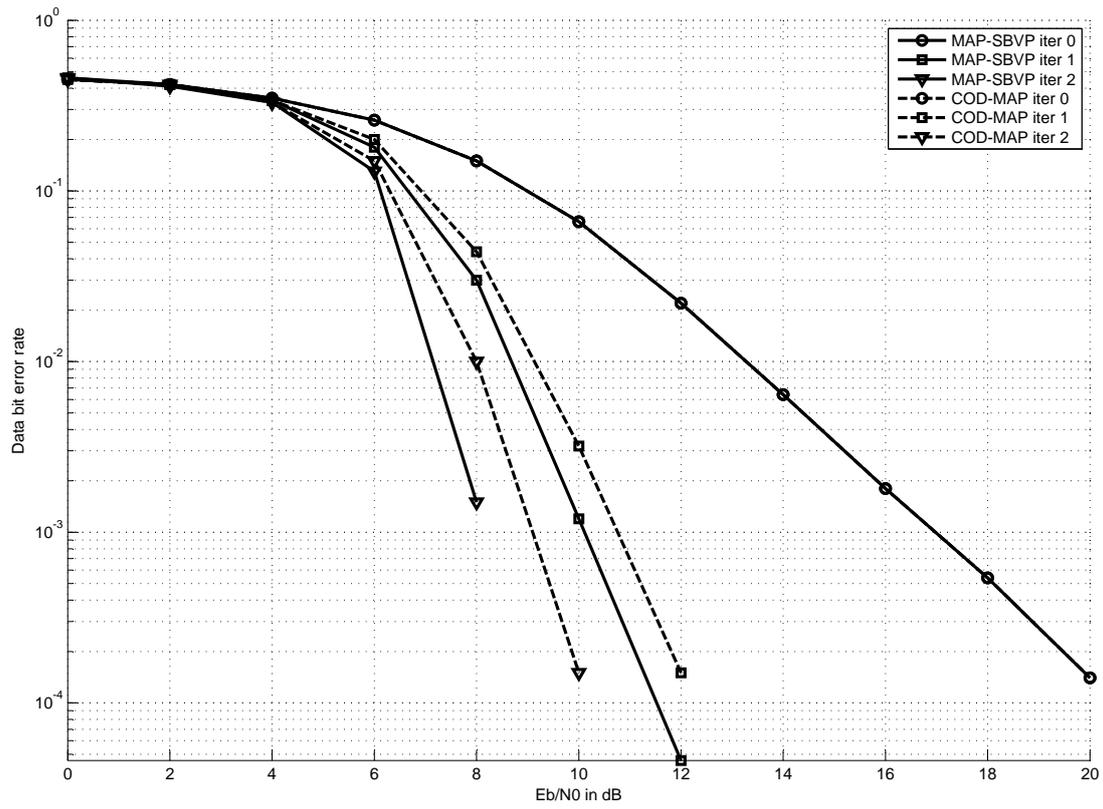

Fig. 4. Channel (b), QPSK Modulation, MMSE Equalization, Rate 1/2 (K=5) Convolution Coding and Puncturing, Block interleaving

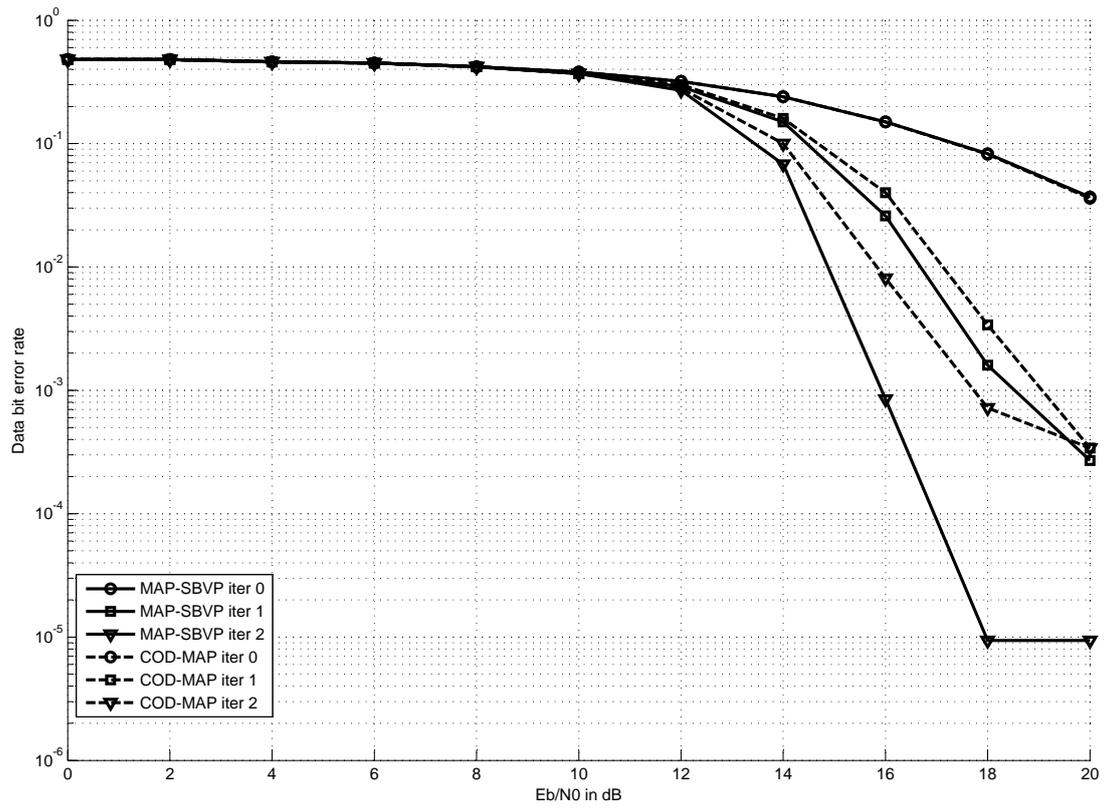

Fig. 5. Channel (c), QPSK Modulation, MMSE Equalization, Rate 1/2 (K=5) Convolution Coding and Puncturing, Block interleaving